\documentclass[manuscript,screen,  ]{acmart}

\setcopyright{acmlicensed}
\copyrightyear{2024}
\acmYear{2024}
\acmDOI{XXXXXXX.XXXXXXX}
\renewcommand\footnotetextcopyrightpermission[1]{} 
\acmISBN{978-1-4503-XXXX-X/18/06}

\usepackage{colortbl}
\usepackage{pifont}
\usepackage{graphicx}
\usepackage{textcomp}
\usepackage{booktabs}

\usepackage{changepage}
\usepackage{algpseudocode}
\usepackage{fancyvrb}
\usepackage{multirow}

\usepackage[utf8]{inputenc}
\usepackage{fancyhdr}
\usepackage{xcolor}
\PassOptionsToPackage{hidelinks}{url}
\PassOptionsToPackage{hidelinks}{hyperref}
\usepackage{wrapfig}
\usepackage{framed}

\makeatletter
 {\par\unskip\endMakeFramed}
\makeatother

\definecolor{formalshade}{rgb}{0.93,0.93,0.93}

\definecolor{darkblue}{rgb}{0.2, 0.2, 0.2}

\newenvironment{formal}{%
  \def\FrameCommand{%
    \hspace{1pt}%
    {\color{darkblue}\vrule width 2pt}%
    {\color{formalshade}\vrule width 4pt}%
    \colorbox{formalshade}%
  }%
  \MakeFramed{\advance\hsize-\width\FrameRestore}%
  \noindent\hspace{-1pt}
  \begin{adjustwidth}{}{}%
  \vspace{2pt}\vspace{2pt}%
}
{%
  \vspace{3pt}\end{adjustwidth}\endMakeFramed%
}

\usepackage{pgfplots}
\pgfplotsset{compat=1.18}
\usetikzlibrary{patterns} 

\usepackage{amsmath,amssymb,amsfonts}

\usepackage{enumitem}
\setlist{noitemsep,topsep=0pt,parsep=0pt,partopsep=0pt}

\definecolor{lightblue}{rgb}{0.85,0.9,1}
\definecolor{blue}{rgb}{0.5,0.75,1}
\definecolor{darkblue}{rgb}{0,0.5,1}

\definecolor{lightred}{rgb}{1,0.85,0.85}
\definecolor{red}{rgb}{1,0.5,0.5}
\definecolor{darkred}{rgb}{1,0,0}

\definecolor{codeblue}{rgb}{0.1,0.1,0.7}
\definecolor{codegreen}{rgb}{0.1,0.6,0.1}
\definecolor{codegray}{rgb}{0.5,0.5,0.5}
\definecolor{codepurple}{rgb}{0.58,0,0.82}
\definecolor{backcolour}{rgb}{0.95,0.95,0.92}
\definecolor{verylightgray}{rgb}{0.95, 0.95, 0.95}
\usepackage{bm}
\makeatletter
\AtBeginDocument{\DeclareMathVersion{bold}
\SetSymbolFont{operators}{bold}{T1}{times}{b}{n}
\SetMathAlphabet{\mathrm}{bold}{T1}{times}{b}{n}
\SetMathAlphabet{\mathit}{bold}{T1}{times}{b}{it}
\SetMathAlphabet{\mathbf}{bold}{T1}{times}{b}{n}
\SetMathAlphabet{\mathtt}{bold}{OT1}{pcr}{b}{n}
\SetSymbolFont{symbols}{bold}{OMS}{cmsy}{b}{n}
\renewcommand\boldmath{\@nomath\boldmath\mathversion{bold}}}
\makeatother

\def\BibTeX{{\rm B\kern-.05em{\sc i\kern-.025em b}\kern-.08em
    T\kern-.1667em\lower.7ex\hbox{E}\kern-.125emX}}

\begin{document}


\title[ LLMs for Optimization]{
Can   Large Language Models Improve SE Active Learning via Warm-Starts?}

 \author{Lohith  Senthilkumar} 
\email{lpanjal@ncsu.edu}
\author{Tim Menzies}
\authornotemark[1]
\orcid{0000-0003-3896-7228}
\email{timm@ieeee.org}
\affiliation{%
  \institution{NC State} 
  \country{USA}
}

\begin{abstract}
When SE data is scarce,  ``active learners''   use   models learned from tiny samples of the data to find the next most informative example to label. In this way, effective models can be generated using very little data.
For multi-objective software engineering (SE) tasks,
active learning can benefit from an effective 
 set of initial guesses (also known as ``warm starts'').  This paper explores the use of Large Language Models (LLMs) for creating warm-starts. Those results are  compared   against   Gaussian Process Models and Tree of Parzen Estimators. For 49 SE  tasks, LLM-generated warm starts significantly improved the performance of low- and medium-dimensional tasks. However, LLM effectiveness  diminishes in high-dimensional problems, where Bayesian methods like Gaussian Process Models perform best.  

In order to better support open science, all the scripts and data used in this study  is available online at \url{https://github.com/timm/moot} and \url{https://github.com/lohithsowmiyan/lazy-llm}
\end{abstract}


 
\newcommand{\bi}{\begin{itemize}}
\newcommand{\ei}{\end{itemize}}

\maketitle

\section{Introduction}
Many problems in SE must trade off between competing constraints;
e.g
\bi
\item
How to deliver {\em more} code but at {\em less} cost?;
\item
How to answer database queries {\em faster} but use 
{\em less} energy?
\ei
Multi-objective  optimization algorithms are tools
that try to satisfy, as far as possible, these competing constraints~\cite{harman2012search},
 Many of these algorithms
begin with an initial set of guesses, then refine them iteratively~\cite{mkaouer2015many,zhang2017constraint}. 
One way to improve such  optimization is to improve the  initial
guess. Such  ``warm start'' policies use some background knowledge to make informed decisions about those  initial guesses.  
For example, warm starts can be generated by asking the opinion of some subject matter expert (SME)~\cite{hacohen2022active, liu2024large, yehuda2022active}.

 As described later in this paper (in \S\ref{slowdata}), SMEs are often in short supply.  
Real-world SMEs are experts precisely because their expertise is valuable to the organization. 
This means that 
  SMEs are often  called away to other tasks. So, where can we find the expertise needed to generate the warm starts?

 Large Language Models (LLMs) have proved useful in many    areas of SE~\cite{hou24,watson22,brown2020language,Tawosi23}.
 Hence it is timely to ask   if    LLMs can  help the optimization of SE tasks  by synthesizing 
 plausible initial guesses.  
As shown by the literature review of this paper, this is an underexplored area, particularly for multi-objective, tabular SE data where labeling  is slow and/or expensive. 
Accordingly, we explore LLMs and warm starts for  the  49 multi-objective SE optimization tasks ~\cite{chen2018sampling, green2009understanding, lustosa2024learning, me07e, menzies2009avoid, nair2016accidental, nair2018finding, port2008using} shown in 
Table~\ref{dataset}.
To structure this investigation, we ask:
\bi
\item {\bf RQ1}: {\em Is   active learning useful for SE tasks?} 
This is our initial baseline question. 
By comparing simple random selection to active learners,
this first questions asks if   active learning is no better
than much simpler methods. 
\item {\bf RQ2}: {\em Are   warm starts   useful for active learning?}
The premise of this work is that the warm start tactic is
worthy of exploration. To check that, this paper compares     active learning results, with and without warm starts. 
\item {\bf RQ3}: {\em Are   LLMs the best way to generate warm starts?} This is main research question of this paper.
Here we will   compare  LLM-driven warm starts to alternative methods.
\ei
As seen in the results of this paper, {\bf RQ1}  will show the superiority of active learning over random methods; {\bf RQ2} will demonstrate the value of warm starts; and {\bf RQ3} will show that, sometimes, LLMs are good way to generate those warm starts.
LLMs  significantly improve outcomes for low- and medium-dimensional problems, reducing the need for extensive labeled data. However, for high-dimensional tasks, LLMs face challenges in synthesizing effective examples, highlighting the continued relevance of Bayesian methods in such scenarios.
 
This study makes the following contributions:
\bi
\item A novel method leveraging LLMs to warm-start active learning for SE optimization tasks (see \S\ref{algo}).
\item An empirical comparison of LLM-based methods with alternate approaches across 49 datasets (see \S\ref{Results}).
\item 
Insights into the strengths and limitations of LLMs in addressing multi-objective SE problems
(see \S\ref{rq3}).
\item
A reproducible package of data and scripts for benchmarking active learning strategies\footnote{ \url{Data: https://github.com/timm/moot}. Code:
  \url{https://github.com/lohithsowmiyan/lazy-llm}}.
\ei
The rest of the paper is organized as follows: Section 2 discusses our motivation and related work, Section 3 details our methodology, Section 4 presents experimental results, and Section 5 discusses implications and future directions.

\begin{table}[!t]
\caption{Data sets  used in this paper. $x/y$ shows the number of independent $x$ vales and dependent $y$ values. The last column show the heuristic categorization of   data sets by
Di Fiore et al.~\cite{difiore2024}  (data sets with  less than 6 or 11 $x$ columns are 
{\em low} or 
{\em medium} complexity while other data sets are
{\em high} complexity).  
For further notes on this data, see \S\ref{data}.}\label{dataset}
 
{\scriptsize
\begin{tabular}{p{1.5cm}|crcl}  
\textbf{Groups} & \textbf{File Name} & |rows| & \textbf{$x$ / $y$} & \textbf{Dims.} \\ \hline
\multirow{32}{*}{Software  Config} 
    & SS-A    & 864    & 3/2  & low      \\
    & SS-B    & 972    & 3/2  & low      \\
    & SS-C    & 1080   & 3/2  & low      \\
    & SS-D    & 2736   & 3/2  & low      \\
    & SS-E    & 3008   & 3/2  & low      \\
    & SS-F    & 3839   & 3/2  & low      \\
    & SS-G    & 5184   & 3/2  & low      \\
    & SS-H    & 4608   & 4/2  & low      \\
    & SS-I    & 6840   & 5/2  & low      \\
    & SS-J    & 65536  & 6/2  & medium   \\
    & SS-K    & 86058  & 6/2  & medium   \\
    & SS-L    & 1023   & 11/2 & low      \\
    & SS-M    & 864    & 17/3 & high     \\
    & SS-N & 86058 & 18/2 & high \\
    & SS-O    & 972    & 11/2 & high     \\
    & SS-P    & 1023   & 11/2 & high     \\
    & SS-Q    & 2736   & 13/3 & high     \\
    & SS-R    & 3008   & 14/2 & high     \\
    & SS-S    & 3840   & 6/2  & medium   \\
    & SS-T    & 5184   & 12/2 & high     \\
    & SS-U    & 4608   & 21/2 & high     \\
    & SS-V    & 6840   & 16/2 & high     \\
    & SS-W    & 65536  & 16/2 & high     \\
    & SS-X    & 86058  & 11/3 & high     \\
    & Apache AllMeasurements & 192 & 9/1 & medium \\
    & SQL AllMeasurements    & 4653 & 38/1 & high \\
    & X264 AllMeasurements   & 1152 & 16/1 & high \\
    & rs-6d-c3 obj1          & 3840 & 6/1  & medium \\
    & rs-6d-c3 obj2          & 3840 & 6/1  & medium \\
    & sol-6d-c2-ob j1        & 2866 & 6/1  & medium \\
    & wc-6d-c1-ob j1         & 2880 & 6/1  & medium \\
    & wc+sol-3d-c4-ob j1     & 196  & 3/1  & low \\
    & wc+rs-3d-c4-obj1       & 196 & 3/1 & low\\
    & wc+wc-3d-c4-ob j1      & 196  & 3/1  & low \\
    \hline
\multirow{10}{*}{\begin{tabular}[c]{@{}c@{}}Software Process \\ Modeling\end{tabular}} 
    & pom3a          & 500    & 9/3 & medium \\
    & pom3b          & 500    & 9/3 & medium \\
    & pom3c          & 500    & 9/3 & medium \\
    & pom3d          & 500    & 9/3 & medium \\
    & xomo\_flight   & 10000  & 23/4 & high \\
    & xomo\_ground   & 10000  & 23/4 & high \\
    & xomo\_osp      & 10000  & 23/4 & high \\
    & xomo\_osp2     & 10000  & 23/4 & high \\
    & coc1000        & 1000   & 17/5 & high \\
    & nasa93dem      & 93     & 22/4 & high \\
    \hline
Project  Health 
    & healthCloseIsses12mths0001-hard & 10000 & 5/1 & low \\
    & healthCloseIsses12mths0011-easy.csv & 10000 & 5/1 & low \\
    \hline
\multirow{3}{*}{Miscellaneous} 
    & auto93         & 398    & 5/3 & low \\
    & Wine\_quality  & 1599   & 10/2 & medium \\
    & HSMGP num      & 3456   & 14/1 & high  
\end{tabular}}
\end{table}

\section{Background}


\subsection{Motivation}
 This research
team has been exploring optimization for SE for many years usually via stochastic methods~\cite{Me07,green2009understanding,chen2017beyond,mathew2017shorter,nair2017flash,DBLP:conf/re/FeatherM02}. Given the random nature of that approach, the question is often asked ``would  a more informed approach result in better and/or faster optimizations?''.

To address that question, at the start of 2024, we  set out to measure the effects of an informed start, on a large number of SE optimization problems. To that end, we  collected   examples from the SE literature where
optimizers had automatically selected    parameters for
(e.g.) cloud services or data miners or software process options. This resulted in   dozens of  tables of data summarized in  Table~\ref{dataset} (and
for full details on that data, see  \S\ref{data}).  
Those tables of data had
\bi
\item
93 to 86,000 rows;
\item
three to 38 independent $x$ variables; and
\item
one to five dependent $y$ variables. 
\ei
In practice, for any row in these tables, the dependent variables values are     unknown until some optimizers decides to test  a particular set of independent values. The process of computing the dependent variables is called 
{\em labeling}.
As discussed in   \S\ref{quirks},   labeling can be very slow and/or expensive. Hence, our   optimizations should achieve their goal using    less than a few dozen labels (for a justification of that labeling budget, see \S\ref{slowdata} and \S\ref{rig}).

\subsection{A Gap in the Literature }\label{gap}
Having assembled test data, and defined the optimization task, the literature was searched for  work trying an informed approach for the better generation of initial candidate   solutions. 
After some initial reading, it was seen that  ``active learning'' and ``warm start'' returned   relevant papers.  
So
in September 2024,   Google Scholar was queried for papers from the last ten years containing those two terms.
The first 1000 papers from that search were sorted
on their citation counts. We observed a knee in that citation curve at   37 citations, above which were 47 papers. Subsequent filtering \footnote{We accepted papers that     mentioned 
  (a)~warm-start used for label initialization; or
  (b)~testing on tabular data; or (c)~mentioned the number of data sets used to evaluate the algorithm; or (d)~mentioned the evaluation budget.}   yielded the papers of Table~\ref{tab:active_learning}, which we summarize as follows:
\bi
\item
{\bf $\ge$ 5 test sets?}: We check our conclusions on 49 data sets. Many other papers Table~\ref{tab:active_learning} use  less than half a dozen.   
\item
{\bf Multi-objective?}: Not only do we explore more data, we explore more complex problems. While 30  of our data sets have multiple goals, most prior work in this area  explored tasks with only a single objective.
\item
{\bf Warm Init?}: Very few papers in Table~\ref{tab:active_learning}   used some background knowledge  to generate warm starts (e.g. an LLM or results from a prior run).
\item
{\bf Data = Tabular?}: Several other researchers also explore tabular data.
\item
{\bf Labels < 50?}: Many papers in Table~\ref{tab:active_learning} gave their algorithms   access  to data from thousands of labeled examples (or even more). However, as discussed in the next section, we have reasons to constrain evaluations to just a few dozen.
\ei
When expressed as a Venn diagram in Table~\ref{tab:active_learning}, this revealed
a gap in the literature (see the center of that diagram).
In summary, unlike this paper, 
 most prior studies (a)~did not handle tabular data; or
 (b)~focused on single-objective tasks where (c)~it was practical to find $10^3$  labels per data set (and sometimes, much more). Note that (c) violates our assumption that we should label only a few dozen examples.

\begin{table}[!t]
  \caption{Literature review results.  Cells marked as ``-'' denotes       no information on that point in that paper. The center of the right-hand-side Venn diagram is the gap
  in the current literature, explored by this paper.}
  \label{tab:active_learning}
  \scriptsize
  \begin{minipage}{3.5in}
  \begin{tabular}{c@{~~~}|r@{~~~}|c@{~~~}|c@{~~~}|c@{~~~}|c@{~~~}|c@{~~~}|c} 
        &          &      &    Warm         & Data=        &     labels        &   $\ge 5$            &Multi- \\
    Ref & Citations & Year &  Init? & Tabular? &  $<50?$ &     test sets? & Objective? \\
    \midrule
    \cite{konyushkova2017learning} & 361 & 2017 & \ding{51} & \ding{51} & - & \ding{55} & \ding{55} \\
    \cite{siddhant2018deep} & 222 & 2018 & \ding{55} & \ding{55} & - & \ding{55} & \ding{55}\\
    \cite{yuan2020cold} & 184 & 2020 & \ding{55} & \ding{55} & \ding{55} & \ding{55}  & \ding{55} \\
    \cite{dong2020mamo} & 163 & 2020 & \ding{55} & \ding{55} & - & - &\ding{55}\\
    \cite{ash2020warm} & 156 & 2020 & \ding{55} & \ding{55} & - & - & \ding{55}\\
    \cite{lowell2018practical} & 122 & 2018 & \ding{51} & \ding{55} & \ding{55} & \ding{51} & \ding{55}\\
    \cite{liu2018learning}& 95 & 2018 & \ding{55} & \ding{55} & \ding{55} & \ding{55} & \ding{55} \\
     \cite{schroder2021revisiting} & 79 & 2021 & \ding{55} & \ding{55} & \ding{55} & \ding{51} &\ding{55} \\
     \cite{hu2018active} & 65 & 2018 & \ding{51} & \ding{51} & \ding{55} & \ding{55} &\ding{55}\\
    \cite{mahmood2021low} & 38 & 2021 & \ding{51} & \ding{55} & \ding{55} & \ding{51} &\ding{55}\\   
    \cite{konyushkova2018discovering} & 36 & 2018 & \ding{55} & \ding{55} & - & \ding{51} &\ding{55} \\
    \cite{chen2024making} & 16 & 2024 & \ding{51} & \ding{55} & \ding{55} & \ding{55} &\ding{55}\\
    \cite{ban2022improved} & 8 & 2022 & \ding{51} & \ding{51} & - & \ding{55} &\ding{55} \\
    \cite{das2023continual} & 2 & 2023 & \ding{55} & \ding{55} & \ding{55} & \ding{51} &\ding{55}\\
    \cite{wei2024basal}& 1 & 2024 & \ding{51} & \ding{55} & \ding{51} & \ding{55} &\ding{55}\\
    \cite{hacohen2022active} & 96 & 2022 & \ding{51} & \ding{55} & \ding{51} & \ding{55} &\ding{55}\\
    \cite{yehuda2022active} & 37 & 2022 & \ding{51} & \ding{55} & \ding{51} & \ding{55} &\ding{55}\\
    \hline
  \rowcolor{green!20} && This Paper & \ding{51} & \ding{51} & \ding{51} & \ding{51} & \ding{51}  
  \end{tabular}
\end{minipage}~~~\begin{minipage}{2.5in}
\includegraphics[width=\linewidth]{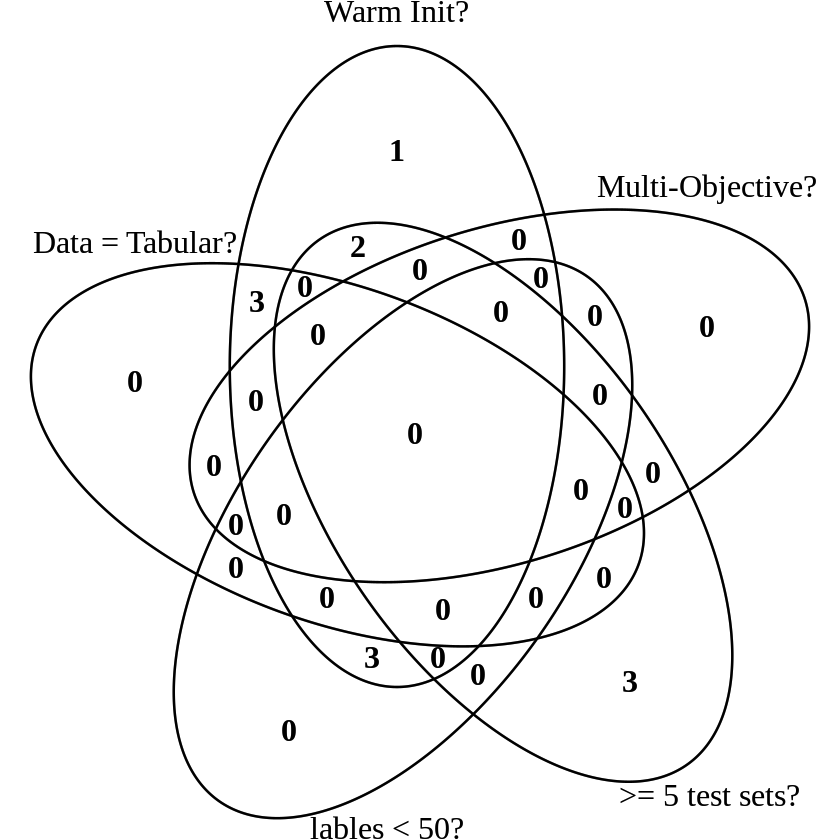}
\end{minipage}
\end{table}

\subsection{The Problem of Not Enough Data}

Why does this paper assume that  we should only evaluate  a few dozen examples? This section argues that this an important  assumption   since there are many problem domains in SE where data is limited. Specifically,
data can be in short supply   for many reasons including
   {\em naive or faulty data} collection (see \S\ref{naive});
     {\em quirks} of  data collection
(see \S\ref{quirks});
  or {\em slow data} collection (see \S\ref{slowdata}).

\subsubsection{ Naive or Faulty  SE Data Collection}\label{naive}
The lesson of decades of SE
analytics  is that even  when   data seems readily available, much of it may be of
dubious quality. 
For example, defect prediction researchers~\cite{catolino2017just, hindle2008large, kamei2012large, kim2008classifying, mockus2000identifying} often  label a commit as "bug-fixing" when the commit text uses words like ``bug, fix, wrong, error, fail, problem, patch''. Vasilescu et al.~\cite{vasilescu2018personnel, vasilescu2015quality} warns that this can be somewhat ad hoc, particularly if researchers just peek at a few results, then tinker with regular expressions to combine a few keywords.
For another example,  Yu et al.~\cite{9226105} explored labels from prior work exploring technical debt, over 90\% of the "false positives" were incorrectly labeled.

More generally, 
there are many reports where errors in data labels have corrupted majority of the examples for security bug labeling~\cite{ 9371393}; for labeling false alarms in static code analysis~\cite{10.1145/3510003.3510214}; or for software code.

When data collect is naive or faulty, some secondary process must clean up the data; e.g.    select a small subset of the data that is not contaminated by quality concerns. 
 If we could reduce the amount of data needed by our learners, then this would reduce how much effort must be allocated to this clean-up process.  Hence, we choose to explore methods that only need a few dozen examples.

\subsubsection{Quirks of SE  Data Collection}\label{quirks}

Another reason why we may lack data on SE problems is a quirk in the nature
of SE data collection. Specifically, in software engineering, the cost  of collecting  independent and dependent information can be very different.
 Consider tabular data where   rows store examples of
 generated from some   function {\em y=f(x)}:
\bi
\item  $x$ columns hold data on independent input observables and/or controllables (e.g.
lines of code)
\item
 $y$ columns hold data on the
 dependent   values (e.g. bugs per line).
 \ei
As stated above, we say that:
\begin{itemize}
\item
{\em Unlabeled} data has $x$ values, with no associated $y$ values
\item
{\em Labeling} means using the $x$ values to find  associated $y$ values;
\end{itemize}
A repeated effect in SE  is that
\bi
\item
We can access copious amounts of unlabeled $x$ data (e.g. from open source projects at Github);
\item
But it is  much harder to collect      high quality labeled $y$ data.
\ei
For example, while it is easy to find software on Github. It is much harder  to determine how much a team spent to create that software.

Table~\ref{slow} lists other examples were
$y$ values are much more expensive to
collect than $x$ values.
 For SE tasks like those of Table~\ref{slow},  
 we need methods that can operate,
 despite a shortage in   training data about the $y$ values.

\begin{table}[!t]
\definecolor{CustomDarkRed}{RGB}{139, 0, 0} 
{\footnotesize
    \centering
    \begin{tabular}{p{2.7in}|p{2.7in}}
   Find $X$ &  Find associated $Y$ values\\
 \rowcolor{CustomDarkRed}    \textcolor{white}{   It can be very  quick to ...} & \textcolor{white}{ It is a much    slower task  to... }\\ 
   
Mine GitHub to find  all the  distributions of code size, number of dependencies per function, etc.
& Discover (a)how much that software  could be sold on the market or (b)  what is the time required
to build this kind of software\\

\rowcolor{gray!20} Count the number of classes in a system. &
Negotiate with an organization permission to find how much human effort was required to build and maintain that code.\\
       List   design options; e.g. 20 binary choices is $2^{20} > 1,000,000$  options.  &  Check all those options with a group of human stakeholders.\\
       
 \rowcolor{gray!20}       List the   configuration parameters for some piece of software. & Generate a separate executable for
       each one of those parameter settings, then run those executable through some test suite.\\    
       
      List the   controls of a data miners used in software analytics  (e.g. how many neighbors to use in a k-th nearest neighbor classifier). & Run a grid search looking for 
the best settings for some local data.\\

 \rowcolor{gray!20} Generate test case inputs (e.g.) using some grammar-based fuzzing.& 
Run all all those tests. It can be even slower for a human to check through all those results
looking for  anomalous behavior.\\
 
    \end{tabular}}
    \caption{For $ y=f(x)$, it is often   cheaper to  collect $x$ values     than the 
associated $y$ values.}
    \label{slow}
\end{table}

\subsubsection{Slow Data Collection from SE Experts}\label{slowdata}

Sometimes, data is missing, since it is fundamentally slow to collect.
Requirements engineering researchers in software engineering report the rate at which subject matter experts (SMEs) can generate  high quality labels; i.e. labels that would be acceptable to a panel  of other experts.   
Numerous sources report that  high quality labels can be collected  from SMEs at the rate of   10-20 items/hour:
\bi
\item 
Valerdi~\cite{valerdi2010heuristics} worked with 
{\em a panel of experts}    labeling  60 software project effort estimation examples, described using   20 attributes. He needed  3*three hour sessions, spread out over a   week. 
\item In iSBSE ({\em interactive search-based software engineering}) humans can serve as one of the oracles to guide the search.  Research on the  SNEAK iSBSE tool~\cite{lustosa2024learning} reported that when SNEAK worked with humans, it collected human insight at   the rates suggested above (15-20 labels per hour).
\item
{\em Repertory grids} researchers conduct   interviews where humans justify attributes and attribute settings for    random subsets of 3 examples, drawn from a larger set. 
    Easterby-Smith~\cite{EASTERBYSMITH19803} advises ``keep the (repertory) grid small. A grid containing ten elements and ten constructs may take two hours to complete. Larger grids may take substantially more time''.
Kington~\cite{kington2009defining} agrees, saying  that it takes humans an hour to reflect over  16 examples with 16 attributes using repertory grids.
\item
In prior work on {\em knowledge acquisition for smart software}~\cite{menzies1999critical,menzies1992expert}, we found that  SME experts may only be  accessible for only a few hours per week. Humans quickly grow exhausted as   they struggle
to explain their expertise (so it is advisable to run   short   knowledge acquisition sessions~\cite{EASTERBYSMITH19803,kington2009defining}). 
\ei
Just to say the obvious, when data is collected at 10 to 20 items per hour (for only a few hours per week), then there will be many cases where new problems  will lack   training data for 
the task at hand. 





\subsection{Methods for Handling Data Shortages}
The data shortage problem in SE has been widely studied.
Various solutions have been proposed including
label-less learning (see \S\ref{less}),
metamorphic learning, semi-supervised learning (see \S\ref{semi}), few-shot learning with LLMs (see \S\ref{llm1}),  and active learning (see \S\ref{active}).


\subsubsection{Label-less Learning}\label{less}
 Zero-shot learning and unsupervised learning are  two classes of algorithms  that can execute, despite a lack of $y$ labels.
 Zero-shot learners use  the  background knowledge of a LLM  to make decisions without needing new labels~\cite{alhoshan2022zero}.
 Zero-shot learners  works in domains where there exists an
  appropriate large language model, which is not always the case.

 Unsupervised learners, on the other hand,  use some domain heuristic to classify examples by 
 peeking at the independent $x$ variables.
 For example,
Nam et al. successfully predicted for defective modules by looking for classes that are unusually large on multiple dimensions~\cite{nam2015clami}.  
For another example, there may be some  general condition
 under which an example
can be unequivocally labeled as ``fail'' such as:
\bi
\item A test case generates a core dump;
\item A metamorphic predicate~\cite{chen2018metamorphic} reports a problem. For an example of such a predicate, consider
  ``small changes to inputs should not cause large changes to 
output''.
\ei
The problem here is that useful domain heuristics    may not be available for all domains. 
 For example, in the case studies shown below, we test for a very wide variety of user-supplied 
 local goals such as ``how to reduce energy requirements'' and ``if we implement this requirement next, then some other team should not stand idle waiting for some other function we were meant to implement''. We are unaware of  (say)  metamorphic predicates that apply to such tasks.

 \subsubsection{Semi-Supervised Learning}\label{semi}
 Yet another approach  finds  labels for just a few examples,  then propagates those values to other near-by examples.
 For example, Yehida et al.~\cite{yedida2023find} recursively bi-clustered $N$ examples down to leaf clusters of size $N^{0.25}$. All examples in a cluster are then given a label computed from the cluster centroid.
 .  
 
Such   semi-supervised learners have successfully reasoned over 10,000s  of  records, after labeling just 1 to 2.5\% percent of the examples~\cite{10109333,majumder2024less}.
While a useful approach, in the studies we have seen~\cite{10109333,majumder2024less},  1 to 2.5\% 
of the data still means  
100s to 1000s  of labels. Given the numbers seen in \S\ref{slowdata} (10 to 20 items per hour, for  few hours per week), we would hope to have methods that work using just a few dozen labels.

\subsubsection{Few-shot Learning with LLMs}\label{llm1}

 In turns out that LLMs need a few examples to activate  the latent space of possible solutions within their networks. Hence, before asking a query, it is best to offer some ``warm-up prompts'' related to the task at hand. 
 In this pre-query process, analysts offer to the LLM a few examples of desired inputs/outputs. 
 Using this approach, a surprisingly small number of examples  (sometimes as few as ten)   
 can   convert a general large  language models into some specific tool; e.g. 
 parsing   test  case output~\cite{le2023log} or translating functions into English~\cite{10.1145/3551349.3559555}. 

One problem with few-shot learning is how to find  a few good examples. Tawosi et al. \cite{tawosi2023search} uses genetic algorithms to select their few shot examples . But that approach has   scalability issues (due to the overheads of the GA).  When we applied it to the 100,000s of examples in our optimization data sets, it  took hours complete the pre-query process.
 Without guidance on
 which example was most informative, these GAs  were taking hours to process all these examples. 
 Hence, for this work, we turned to methods that   pruned away most of the examples before collecting labels. Specifically, we use active learning. In stark contrast to the Tawosi et al. \cite{tawosi2023search} method, our approach can terminate in just a few sections (in particular, the ``explore'' and ``exploit'' TPE methods discussed below).

\subsubsection{Active Learning}\label{active}

Active learning
is a technique that (a)~reflects on a model built so far to (b)~determine which examples to
label next.  
A repeated result is that this approach learns good models, using very little data~\cite{brochu2010tutorial}.

 Active learner can begin via a {\em cold start} that   label a few examples  via a random or unsupervised process; e.g.
 
\begin{enumerate}[]
\item {{\em Random sampling}}: This approach just selects   items at random. While certainly the simplest and fastest method, the experiments of this paper show that other approaches give better results.

 \item {{\em Similarity sampling}}: Similarity sampling selects subsets of the examples where, within each subset, there is much similarity~\cite{lin2023similarity}.

\item {{\em Diversity sampling}}: Diversity sampling selects data points that are different from those already chosen or labeled~\cite{wang2021preference}.  Many   diversity   sampling methods are clustering based \cite{hacohen2022active, yehuda2022active} while other methods like \cite{brinker2003incorporating} introduces diversity sampling methods using maximum distant points from the projections of the hyper planes.

\end{enumerate} 
 There is some evidence that 
 better results come from  {\em warm starts}
 that use some background knowledge to select the initial labels (see \cite{brochu2010tutorial,lustosa2024learning} and the
experiments of this paper).
Methods for generating   warm starts include  (a)~reusing the results from a prior session; or (b) using the knowledge inside an LLMs are select the warm start examples.

 Once some examples have been labeled, then some {\em acquisition function} is applied to decide where to sample next. 
\cite{watanabe2022speeding} claim that Gaussian Process Models  (see \S\ref{GPM}) are less
efficient that Tree of Parzen Estimators
 (see \S\ref{TPE})      for active learning and multi-objective optimization. We will explore both approaches

  \subsubsection{GPM = Gaussian Process Models}\label{GPM}
  GPM  compute the mean $\mu$ and standard deviation $\sigma$
of an estimate by fitting the data to a wide range of possible functions. These $\mu,\sigma$ values are  then used by acquisition functions to decide where to sample next~\cite{williams1995gaussian,brochu2010tutorial}.

 Table~\ref{algogpm} shows the connection of active learning, acquisition functions, and GPM.   This paper explores three GPM-based  acquisition functions: {\em UCB, PI} and {\em EI}.
 
 \begin{table}[!b]
 \caption{Active Learning with Gaussian Process Models}\label{algogpm}\small
 ~\hrule~ 
\begin{enumerate}
\item Evaluate the labels of  some rows  (using a cold or warm start);
\item Fit a large space of possible functions to the labeled data.
\item
From that function space, estimate the mean
$mu(x)$ and standard deviation $\sigma(x)$
for different $x$ values;
\item 
Using those estimates,
apply an acquisition function
to select the row with the most promising $x$ values to label.
\item Evaluate a label at that $x$ point;
 \item If evaluation budget exhausted, return row with best label.
Else, go to step 2.
\end{enumerate}
~\hrule~
\end{table}

 {\textbf{UCB = Upper Confidence Bound}}:
Introduced in 1992 by \cite{cox1992statistical}, UCB is an adaptive acquisition function that recommends sampling the next example that maximizes;
\begin{equation}
    UCB(x) = \mu(x) + \kappa\sigma(x)
\end{equation}

\begin{itemize}
    \item $\mu$ is the   mean predicted at point x.
   
    \item $\sigma(x)$ is the predicted standard deviation at point x.
    \item $\kappa$ is the parameter that balances the exploration and exploitation.
    
\end{itemize}
As active learning progresses, the variance in the predictions decreases so $\kappa\sigma(x)$ term grows smaller w.r.t. the $\mu(x)$ term.
This means that UCB is {\em adaptive} to 
the amount of data collected;
i.e. initially, UCB explores regions of high variance but, as data collection continues, it adapts to exploit regions of best predictions.

  {\textbf{PI = Probability Improvement}}:
Introduced first by \cite{kushner1964new}, Probability Improvement (PI) is an acquisition function used to approximate the best configuration over the parameter space of a noisy distribution. PI calculates the probability of improvement the next incumbent sample brings in to the distribution. 
\begin{equation}
    PI(x) = \phi \left(\frac{\mu(x) - f(x^*) - \epsilon}{\sigma(x)}\right)
\end{equation}

\begin{itemize}
    \item $\mu$ is the predicted mean at point x.
    \item $f(x^*)$ is the current best point
    \item $\epsilon$ is the  parameter balancing explore and exploit.
    \item $\sigma(x)$ is the predicted standard deviation at point x.
    \item $\phi$ is the PDF of the distribution.
\end{itemize}
 {\textbf{EI = Expected Improvement}}
(introduced in \cite{jones1998efficient}):  Expected improvement   measures the expected gain from sampling a new point based on the priors. 
\[
    EI(x) = 
    \begin{cases} 
      (\mu(x) - f(x^*) - \epsilon)\Phi(Z) + \sigma(x)\phi(Z), & \text{if } \sigma(x) > 0, \\
      0, & \text{if } \sigma(x) = 0.
    \end{cases}
\]

\[
    Z = 
    \begin{cases} 
      \frac{\mu(x) - f(x^*) - \epsilon}{\sigma(x)}, & \text{if } \sigma(x) > 0, \\
      0, & \text{if } \sigma(x) = 0.
    \end{cases}
\]

\begin{itemize}
    \item $\mu$ is the predicted mean at point x.
    \item $f(x^*)$ is the current best point
    \item $\epsilon$ is the  parameter balancing explore and exploit.
    \item $\Phi(Z)$ is the Cumulative Distribution Function (CDF) of the standard normal distribution
    \item $\sigma(x)$ is the predicted standard deviation at point x.
    \item $\phi(Z)$ is the PDF of the standard normal distribution.
    
\end{itemize}

\begin{wrapfigure}{r}{1.5in}   
\begin{tikzpicture}[scale=0.67]
    \begin{axis}[
        axis equal image,
        xlabel={$B=best$},
        ylabel={$R=rest$},
        xmin=0, xmax=1,
        ymin=0, ymax=1,
        grid=major,
        axis lines=middle,
        enlargelimits
    ]
        \addplot[dashed, domain=0:1] {x};

        \node[above left] (label) at (axis cs:0.5,0.7) {$B - R \approx 0$};
        \draw[->, thick] (label) -- (axis cs:0.5,0.5);

        \addplot[draw=none, fill=gray!20] coordinates {
            (0.6, 0.6) 
            (1, 0.6) 
            (1, 1) 
            (0.6, 1)
        } \closedcycle;

        \node[above right] (regionLabel) at (axis cs:0.4,0.9) {$B + R \text{ is large}$};
        \draw[->, thick] (regionLabel) -- (axis cs:0.8,0.8);

        \addplot[draw=none, fill=gray!20] coordinates {
            (0.6, 0)  
            (1, 0) 
            (1, 0.4)
            (0.6, 0.4)
        } \closedcycle;

        \node[below right] (bGrLabel) at (axis cs:0.3,0.3) {$B / R $ is large};
        \draw[->, thick] (bGrLabel) -- (axis cs:0.8,0.2);
    \end{axis}
\end{tikzpicture} 
\caption{Our TPE  active learner explores the decision space of a two-class classifier. }\label{br}
\end{wrapfigure}
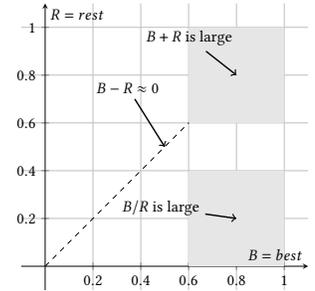
Expected improvement is a adaptive acquisition function that balances between exploration and exploitation, exploration is high when the surrogate variance is higher and exploitation is high when the surrogate mean is higher.

 \subsubsection{TPE = Tree of Parzen Estimators}\label{TPE}

One alterative to using
 Gaussian process models
 are acquisition functions
 that use  Tree of Parzen Estimators (TPE) \cite{bergstra2011algorithms, ozaki2020multiobjective}.
 While Gaussian process-based approaches model \( p(y|x) \) directly (i.e., \( y = f(x) \)), TPE \cite{bergstra2011algorithms} models \( p(x|y) \) and \( p(y) \) separately. 
To do this, 
TPE   divides the sorted observations \( D = \langle x, f(x) \rangle \) according to the label   values into two sets using a threshold \( y^* \). Thus, TPE defines \( p(x|y) \) using two distributions:

\[
p(x|y) =
\begin{cases} 
    best(x) & \text{if } y < y^*, \\
    rest(x) & \text{if } y \geq y^*,
\end{cases}
\]
where \( \mathit{best}(x) \) is the model formed by observations that performed well and \( \mathit{rest}(x) \) is the model formed by observations that performed poorly. As shown in Figure~\ref{br}, these two models let us explore several regions of interest.
For example:
\bi
\item
Where $B - R \approx 0$, it is unclear if some new example is ``best'' or ``rest''.
Some acquisition functions target this ``zone of uncertainty'' since it is here we can learn where an example flips from one class to another.
\item
Where $B/R$ is
large, our two models are in agreement that something is probably ``best''. Some acquisition functions target this ``zone of certainty'' since it is here we can be most certain that some is ``best''.
\item
Where $B - R \approx 0$ and $B+R$ is large. Here, our two models are strongly arguing for different conclusions. 
Some acquisition functions target this ``zone of dispute'' since it is here we can most
learn what features lead to very different conclusions. 
\ei
 \begin{table}[!bt]
 \caption{Active Learning with Tree of Parzen Estimators}\label{algotpe}
 \small
 ~\hrule~ 
\begin{enumerate}
\item Evaluate the labels of some rows (using a cold or warm start);
\item Sort the labeled rows into a small {\em best} set and a larger {\em rest} set.
\item Using that binary division,
build a classifier that   reports the likelihood $B,R$ of an   example being  {\em best} or {\em rest}.
\item 
Using $B,R$
apply an acquisition function
to select the row with the most promising $x$ value to label.
\item Evaluate a row at that $x$ point;
\item If evaluation budget exhausted, return row with best label.
Else, go to step 2.
\end{enumerate}
~\hrule~
\end{table}

 Table~\ref{algotpe}, shows the connection of active learning and acquisition functions. This paper explores two
TPE-based 
acquisition functions:
{\em explore}
and {\em exploit}.

 {\bf{Explore}}: Explore  favors the zone of dispute; i.e.  where two models are both loudly proclaiming similar strength, but opposite, prediction. When examples fall close to the hyperspace boundary that separates examples, 
that are most uncertain with respect to currently available data.   Our formula for Explore is given by:
\begin{equation}
 Explore = \frac{abs(B + R)}{(B - R) + \epsilon}
\end{equation}
Here, $\epsilon$ is a very small number added to avoid divide-by-zero errors.

{\bf{Exploit}}: Exploit   targets the zone of certainty; i.e. where the best and rest models are in agreement that an example belongs to     ``best''.   The formula for Exploit is given by:  
\begin{equation}
    Exploit = \frac{B}{R}
\end{equation}

 \section{Methods}
The experiments of this paper use the data of \S\ref{data} to compare the performance of LLM-aided warm start method  for
\bi
\item
The state of the art GPM methods using the UCB, PI and EI
acquisition functions. 
\item
Against the TPE  based acquisition functions Explore, Exploit;

\ei
This conclusions of this paper were achieved using the methods described in this section.

\begin{table}[!b]
\caption{Overview of MOOT's data.
For notes on the structure of these data sets, see
Table~\ref{dataset}.}
\label{mootoverview}
\small
\centering
\begin{tabular}{|>{\raggedright\arraybackslash}m{1.5in}|>{\raggedright\arraybackslash}m{2in}|>{\raggedright\arraybackslash}m{2in}|}
\hline
\textbf{Dataset Type} & \textbf{Features} & \textbf{Primary Objective} \\ \hline
SS-Models & Runtimes, query times, and usage data from software configured in various manners, selected randomly & Identify configurations that optimally meet the software goals of each project \\ \hline
Specific Software Configurations & Configurations of specific software systems—Apache servers, Hazardous Software Management Growth Program (HSMGP), SQL databases, and X264 video encoding & Optimize particular settings that enhance the performance and efficiency of these well-known applications \\ \hline
General SS-Datasets (RS, SOL, WC) & Runtimes, query times, and usage data from software configured in various manners, selected randomly & Collect various performance metrics from randomly configured software systems tailored toward specific software engineering challenges \\ \hline
Health Data Sets & Random forest regression to predict future GitHub project metrics such as commits, closed issues, and pull requests & Predictive measure of project health and developer activity over a 12-month period \\ \hline
Debug Datasets (Auto93, Wine Quality) & Generalizable data for software engineering newcomers & Auto93 optimizes for car performance metrics, whereas Wine Quality focuses on the attributes influencing the quality of wine \\ \hline
COC1000 and NASA93 & Based on the Constructive Cost Model (COCOMO) approach & Emphasizes the reduction of risk, effort \\ \hline
POM3 (A-D) & Agile project management dynamics & Simulate scenarios where team idle rates, task completion rates, and overall costs are in flux \\ \hline
XOMO (Flight, Ground, OSP, OSP2) & Software process optimizations at NASA’s Jet Propulsion Laboratory & Addresses the complexity and challenges in aerospace software development projects \\ \hline
\end{tabular}

\end{table}

\subsection{Data}\label{data}
This paper applies the acquisition functions described above to data shown in Table~\ref{dataset} from  the MOOT repository
(\url{https://github.com/timm/moot}).
MOOT, short  Multi Objective Optimization Testing is a collection of
  49    optimization/configuration problems taken from the SE multi-objective optimization literature~\cite{chen2018sampling, green2009understanding, lustosa2024learning, me07e, menzies2009avoid, nair2016accidental, nair2018finding, port2008using}.
  For some details on  the MOOT data sets, see Table~\ref{mootoverview}.

  MOOT stores tabular data. Each row contain many $x$ columns (for   input independent variables) and one or more $y$ columns
  (for the output dependent variables). As seen in Table~\ref{dataset}, most   
  of MOOT's data has $|y|>1$; i.e. they are multi-objective problems.  
  To the best of our knowledge, our use of MOOT makes this paper
one of the largest multi-objective SE optimization results
reported in the literature. Having worked in this field since 2002~\cite{DBLP:conf/re/FeatherM02}, we can assert that resources like MOOT are very rare.
  As shown in our literature review (\S\ref{gap}), most research papers in this field certify their methods using fare less data than the 49 current entries within MOOT (usually papers use five data sets, or less). 
  
  MOOT's data is mostly    generated by exercising some SE model (e.g. the ``pom*'' models discussed below)
  or logs of the the operation of project artifacts. As an example of the latter:
  \bi
  \item
  All the ``SS-*'' data sets were generated by mutating the control parameters of a Makefile, then building some project, then running a test suite through that piece of software. 
  \ei

  Note that MOOT stores data rather than models. This was a deliberate choice made for reproducibility reasons. Two different research teams working on (e.g.) ``SS-A'' will be exploring exactly the same set of values. The same is not true if each researcher codes up their own version of some model like (e.g.) DTLZ3~\cite{1007032} then explores it using different random number generators.

  For the purposes of experimentation, all the rows in MOOT's data shows values for the $y$ columns. But recalling 
\S\ref{quirks}, all our experiments will assume that accessing a row's $y$ values is a very expensive process.
  Hence, our algorithms will strive to find the best rows in these data sets after looking at as few $y$ values as possible. 

  Prior work by Di Fiore et al.~\cite{difiore2024} has argued that active learning experimental results should be divided according to the number of  independent variables seen in each optimization problem. Accordingly,
Di Fiore et al.,  categorize  datasets into 3 categories based on their input ($x$) dimensionality:
\begin{itemize}
    \item \textit{Low:} 12 datasets with $|x|<6$ independent features.
    \item \textit{Medium:} 14 datasets with $6 \le|x| \le 11$ independent features.
    \item \textit{High:} 19 datasets with $|x|>11$ independent features.
\end{itemize}
Other papers define ``high dimensionality'' in different ways to the above. For example,  
a common distinction is to refer to  text mining as a lower dimensional problem than image processing. Yet this paper  would call both ``high dimensional''. 
But in defense of our definitions, we note that this paper 
(as well as Di Fiore et al.) find
important differences  in the behavior of acquisition functions over
data sets divided according to the {\em low, medium, high}
definitions listed above.

As to the application areas of MOOT,
it is  divided into four subcategories:
\begin{enumerate}
\item   Datasets under the {\em  Config} directory included datasets with the group names "SS-*" that comes from the software engineering literature \cite{nair2016accidental}.This data was exhaustive collection of running configuration of different tasks including video encoding with the goals mostly being (Query time, Run times, etc). Config also includes other datasets with database configurations including  Apache\_AllMeasurements.csv, SQL\_ALLMeasurements.csv, X264\_AllMeasurements.csv. 
\item
The {\em HPO} directory contains the datasets from the hyperparameter optimization literature \cite{lustosa2024learning} . These datasets show the results of random forest regression algorithms trained to predict a) commits, b) closed issues c) close pull requests in 12 months time on Open source projects hosted at Github. The Y values of these datasets shows the error and accuracy of the model for different hyperparameter configurations. 
\item 
The {\em Process} datasets comes from the software process modeling literature \cite{green2009understanding, Me07, menzies2009avoid, port2008using}. The data set named "pom*" shows data from agile development by \cite{boehm2004balancing}. POM3 models requirements as a tree of dependencies that 
(metaphorically) rises out of pool of water. At any time, developers can only see the dependencies above the waterline. Hence they can be surprised by unexpected dependencies that emerge at a later time. POM3 reports the completion rates, idle times, and development effort seen when teams try navigate this space of changing tasks. The "nasdem" dataset contains real world data and "osp2" show data in the format of the USC Cocomo models that predict for development effort, defects, risk in waterfall- style software projects \cite{Me07}.
\item
The {\em Misc} directory contains some non SE datasets (auto93 and WineQuality). We use these to demonstration MOOT to a non-SE audience. 
\end{enumerate}




\subsection{Models}

Selecting the most suitable large language model for a given task requires careful consideration of various factors. Among the multitude of publicly available models, the most prominent families include GPT, Gemini, and Claude. While prior research \cite{siam2024programming} indicates that GPT outperforms its competitors in software engineering benchmarks. However, our specific task requires a deeper contextual understanding of the dimensions and configurations of the data.

Previous studies \cite{hochmair2024correctness} suggest that Gemini models excel in precise factual reasoning, particularly for tasks demanding high contextual comprehension. Furthermore, Gemini offers a 1-million-token context window \cite{team2024gemini}, significantly surpassing GPT-4o’s 128k tokens. This extended context window allows the inclusion of extensive metadata descriptions and examples, which is especially advantageous for high-dimensional datasets. Additionally, Gemini’s competitive per-token pricing compared to GPT was a decisive factor.

For all the reasons pf the last paragraph, this study used  Gemini 1.5 Pro.
 
\subsection{Algorithms}\label{algo}

   \begin{figure}[!t]
    \centering\footnotesize
     \includegraphics[width=.7\linewidth]{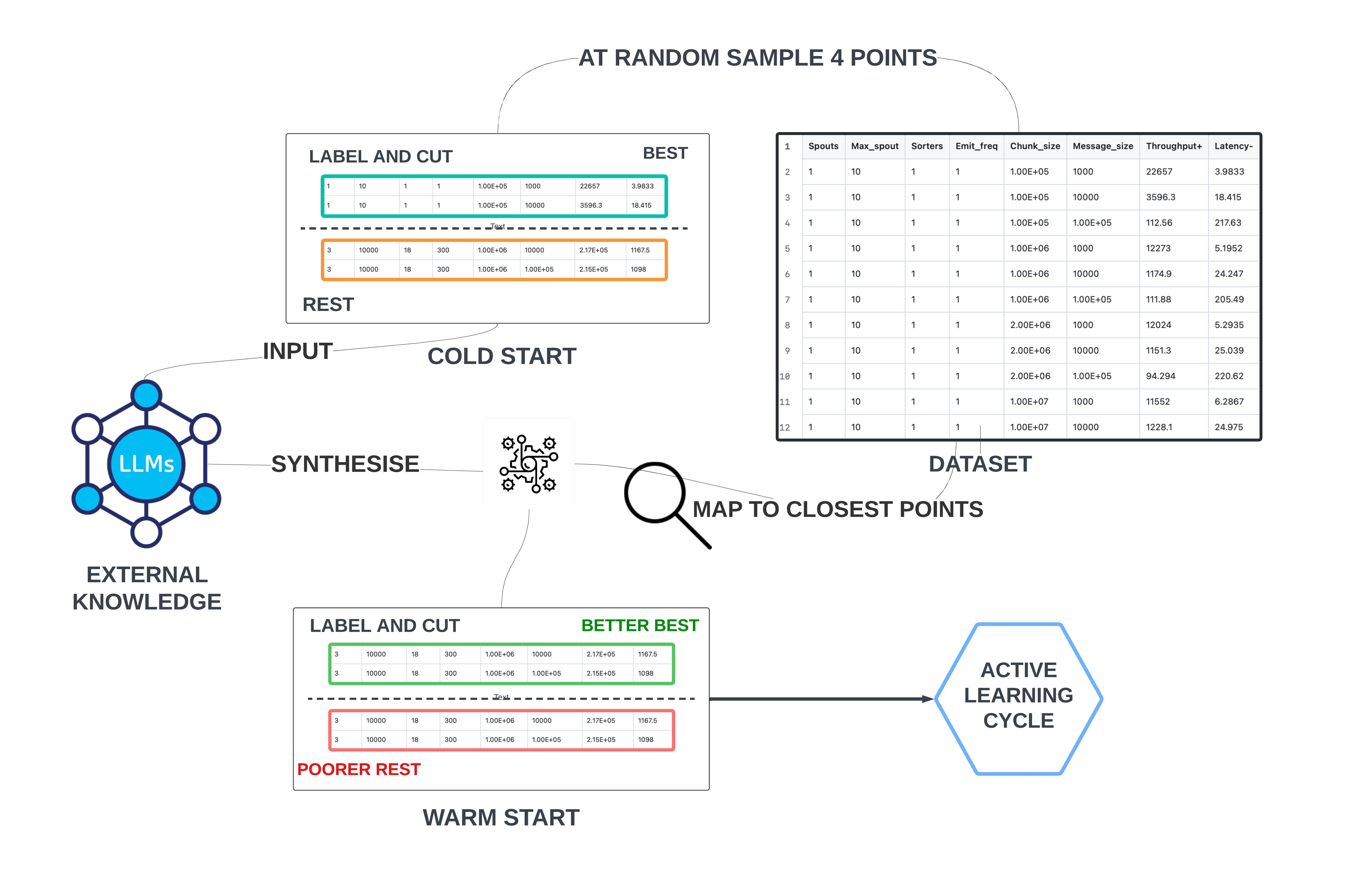}
    \caption{Warm-starting with LLM Synthesized examples in cycle 0 of active learning}
    \label{fig:enter-label}
\end{figure}
\begin{table}[!t]
\centering\footnotesize

\caption{Prompt template for synthetic data generation, here the two variables are meta which is a table containing the meta data of the dataset and table is set of rows that were randomly picked at the zero'th iteration.}\label{tbl:fsl}
\begin{tabular}{|p{0.95\linewidth}|}
\hline

\rowcolor{blue!10}
\textcolor{blue}{\textbf{System Message:}} \\
\texttt{You are given a dataset with several features. The rows have been categorized into \textbf{"Best"} and \textbf{"Rest"} examples based on their overall performance. Below are the key features and their descriptions from the dataset:}\\
... \\
\textcolor{teal!90!black}{\texttt{\{rows\_to\_markdown(meta)\}}} \\[1em]

\hline

\rowcolor{yellow!20}
\textcolor{orange}{\textbf{Human Message:}} \\
\texttt{Given Examples:}\\[1em]
\textcolor{teal!90!black}{\texttt{\{rows\_to\_markdown(table)\}}} \\[1em]

\hline

\rowcolor{pink!20} 
\textcolor{red!70!black}{\textbf{Task:}} \\
\texttt{1. Generate Two New Examples that are Better:} \\
\texttt{These should outperform the given \textbf{"Best"} examples by optimizing the relevant features to better combinations.} \\[1em]

\texttt{2. Generate Two New Examples that are Poorer:} \\
\texttt{These should under perform the given \textbf{"Rest"} examples by modifying the relevant features to worse combinations.} \\[1em]

\texttt{Consider the inter-dependencies between features, and ensure that the generated examples follow logical consistency within the dataset's context.} \\[1em]

\texttt{Return the output in the same markdown structure:} \\

\hline
\label{prompt}
\end{tabular}
\end{table}

\subsubsection{Active Learners}
The active learners used in this study were described in
 Tables~\ref{algogpm} and \ref{algotpe}.

 \subsubsection{Acquisition Functions}
 All the   acquisition
 functions used in this work were described in 
 \S\ref{GPM} and \S\ref{TPE}.

 \subsubsection{Warm Starts with LLMs}
 LLMs were used to generate warm starts as per the method illustrated in 
 Figure~\ref{fig:enter-label}:
 \begin{enumerate}
 \item 
Initial labels were assigned to rows selected  at random.
\item 
These labeled rows, called $E_0$ were then sorted best to rest using the multi-objective evaluation criteria of \S\ref{eval}.
\item
Few-shot learning~\cite{sui2024table} was then applied.
The $E_0$ examples were given as prompts to the LLM, with comments saying "this is a good example", "this is a bad example".
\item
The LLM was then asked to generate the $x$ values of new examples, called $E_1$ that were, in the opinion of the LLM,   better or worse that all the $E_0$ examples.
\item 
Since all the items in $E_1$ were invented by the LLM,
they  have no $y$ column labels. To generate labeled examples, we then went back to the training data and  for each row $r \in E_1$, we found its nearest neighbor from $E_0$ (where ``near'' was measured using the Euclidean distance of the  independent $x$ column values). This formed the
set $E_2 \subset E_0$.
\item
$E_2$ was then used  to warm-start the active learners.
\end{enumerate}
Table~\ref{tbl:fsl} offers more details of the prompting used in few-short learner used in step \#3.
All our prompts, are provide in markdown format for easier reading by the LLM   (a technique recommended by~\cite{sui2024table}).
Note that
 our prompt template was divided 
 into three  parts:
 \begin{itemize}
     \item {\em System message}: where we define the role of our LLM which is a general prompting strategy, 
     along with the meta data, where we provide information about the individual attributes of the dataset including, type, median value, standard deviation, high \& low values if NUM or mode and frequencies if SYM. 
     Note that in this section,   we mention attribute names.
     This has the effect of ``waking up''  all the LLM's background knowledge of those terms.
     \item {\em Few Shot Examples}: Random samples from the data with the features along with their labels (Best / Rest).
     \item {\em Task}: in this case we ask the LLM to generate 2 better and 2 poorer samples while maintaining the same markdown format for better post-processing.
 \end{itemize}

\subsection{Evaluation Criteria}\label{eval}

The above methods need some way to recognize  
  ``best'' and ``rest'' rows.
Following the precedent set by the multi-objective research community~\cite{zhang2007moea}, our rows are  judged best or rest using the Chebyshev distance~\cite{giagkiozis2015methods} between a row's y-values and the theoretical ideal y-values:
  \bi
  \item
  For dependent attributes that we wish to minimum or maximize, the {\em ideal}   y-values contain the minimum or maximum vale (respectively) for that column. 
  \item
  The Chebyshev distance is the maximum distance between any $y_i$ value and its the ideal  $l_i$ value:
  \[
d_\text{Chebyshev}(y, o) = \max_{i=1, \ldots, n} \left| y_i - l_i \right|
\]
\item 
Note that for Chebyshev,
{\em smaller} scores are {\em better} since this means that the row is closer to the ideal values.
Hence, given a set of $N$ labeled row, we sort them ascending via Chebyshev, then declare the first $\sqrt{N}$ rows to be ``best'' and
the remaining
$N-\sqrt{N}$ rows to be ``rest''.
\ei
We use Chebyshev since
there is precedence for its use in the multi-objective research literature~\cite{zhang2007moea,giagkiozis2015methods}. Also it is a ``cruel critic''; i.e. it punishes and demotes a row if any of its $y$ values are bad, even in all the others are performing admirably.

\subsection{Statistical Methods}\label{stats}
We rank our results using the Scott-Knott method \cite{scott1974cluster}. The Scott-Knott method is a recursive bi-clustering approach. It recursively divides the treatments (if they are statistically distinguishable from one other) to  returns with a list of sets of treatments each ranked from [0..n]. 
 Scott-Knott   is recommended over other multiple comparison tests, since it does not produce overlapping groups like other post-hoc tests (e.g., Nemenyi’s test).
Also, Scott-Knott tests for both effect size and significance
(using  the Cliff's Delta and bootstrapping methods described below); it works well with overlapping distributions; N samples are ranked with just \textit{$log_2(N)$} statistical comparisons.

In this work, all treatments (acquisition methods, budget) are sorted in order of their median Chebyshev distances.
Scott-Knott then  divides the list into two halves at the point where there is maximum expected difference between the mean before and after division.

\begin{equation}
    E(\Delta) = \frac{|l_1|}{l} abs(E(l_1) - E(l))^2 + \frac{|l_2|}{l} abs(E(l_2) - E(l))^2 +
\end{equation}
Here. |$l_1$| is the size of list 1 and |$l_2$|is the size of list 2

\begin{wraptable}{r}{3.5in}
\scriptsize
\centering
\caption{Scott-Knott Rankings of the SS-A dataset along with treatments and budgets.}\label{eg}
\begin{tabular}{c|cc|c|cc|l}
     &       & Evaluation &\multicolumn{2}{|c|}{Chebyshev}& Visualization: \\
Rank & Start & Acquire & Budget & \multicolumn{1}{|c}{Median} & \multicolumn{1}{c|}{Std.} & ``o'' = median \\ 
\hline
\rowcolor[HTML]{D3D3D3} 
0 & LLM & exploit & 20 & 0.07 & 0.01 & o-             \\ 
\rowcolor[HTML]{D3D3D3} 
0 & LLM & exploit & 15 & 0.07 & 0.02 & o-             \\ 
\rowcolor[HTML]{D3D3D3} 
0 & random &exploit     & 25 & 0.08 & 0.03 &    o ------       \\ 
\rowcolor[HTML]{D3D3D3} 
0 & LLM & exploit & 25 & 0.08 & 0.01 & o-             \\ 
\rowcolor[HTML]{D3D3D3} 
0 & random & exploit     & 20 & 0.08 & 0.03 & - o--          \\ 
\rowcolor[HTML]{D3D3D3} 
0 & random & exploit     & 15 & 0.09 & 0.03 & - o -- 
\\  
1 &random & PI\_GPM     & 20 & 0.09 & 0.00 & \hspace{2mm}o 
\\  
\rowcolor[HTML]{D3D3D3} 
2 &random & UCB\_GPM    & 25 & 0.09 & 0.00 & \hspace{2mm}o  
\\ 
3 & LLM & explore & 20 & 0.10 & 0.02 &\hspace{1mm}- o- 
\\  
\rowcolor[HTML]{D3D3D3} 
4 & LLM & explore & 25 & 0.10 & 0.05 & \hspace{1mm}- o ---------  \\ 
\rowcolor[HTML]{D3D3D3} 
4 &   &random    & 20 & 0.10 & 0.02 & \hspace{3mm}o-             \\ 
\rowcolor[HTML]{D3D3D3} 
4 & random &EI\_GPM     & 15 & 0.10 & 0.00 & \hspace{3mm}o   
\\  
5 &  & random      & 25 & 0.11 & 0.02 & \hspace{2mm}-- o-          \\ 
5 &  & random      & 15 & 0.12 & 0.03 & \hspace{2mm}-- o           \\ 
5 & LLM & explore & 15 & 0.12 & 0.05 & \hspace{2mm}- o ---------- \\ 
5 & random & PI\_GPM     & 25 & 0.12 & 0.00 & \hspace{4mm}o              \\ 
5 &random & explore     & 15 & 0.13 & 0.04 & \hspace{2mm}- o-------     \\  
\rowcolor[HTML]{D3D3D3} 
6 & random & UCB\_GPM    & 20 & 0.13 & 0.00 & \hspace{4mm}o              \\ 
\rowcolor[HTML]{D3D3D3} 
6 & random & explore     & 20 & 0.13 & 0.04 & \hspace{3mm}- o----------  \\ 
\rowcolor[HTML]{D3D3D3} 
6 &random & UCB\_GPM    & 15 & 0.13 & 0.00 & \hspace{4mm}o              \\ 
\rowcolor[HTML]{D3D3D3} 
6 & random & EI\_GPM     & 25 & 0.13 & 0.00 & \hspace{4mm}o              \\ 
\rowcolor[HTML]{D3D3D3} 
6 &random & PI\_GPM     & 15 & 0.13 & 0.00 & \hspace{4mm}o              \\ 
\rowcolor[HTML]{D3D3D3} 
6 &random & explore     & 25 & 0.14 & 0.04 & \hspace{4mm}- o---------   \\  
7 & & baseline    & 1512 & 0.18 & 0.09 & \hspace{7mm}-- o    --    \\  
\rowcolor[HTML]{D3D3D3} 
8 &random & EI\_GPM     & 20 & 0.36 & 0.00 & \hspace{15mm}o              \\ 
 
\end{tabular}
\label{tab:model_summary}
\end{wraptable}

~{\em Cliff's Delta} is a non-parametric effect size measure that quantifies the difference between two distributions by calculating the proportion of pairwise comparisons. It evaluates how often values from one distribution are larger or smaller than those from another. The result ranges from -1 to 1, where values closer to 0 indicate little difference, and values near -1 or 1 signify strong separation between groups. Cliff’s Delta does not assume normality, making it ideal for non-parametric data. A threshold is used to classify the strength of the observed effect (small, medium, or large).

~{\em Bootstrapping} is a resampling method. By repeatedly drawing samples with replacement from the original data, bootstrapping creates distributions of statistics (like means or medians) to approximate confidence intervals. This technique tests if the observed data is significantly different from random variation, making it useful for hypothesis testing and validating model performance, especially with limited data.

If the results of these statistical tests confirms distinguishably of the groups then Scott-Knott ranks these groups and recursively repeats the process for both these clusters. 

 Table \ref{tab:model_summary} shows the output of the recursive bi-clustering ranking procedure of the Scott-Knott Method with treatments under the statistical rank 0 being the best of performing treatment for the dataset.

 In 
Table~\ref{tab:model_summary}, the statistical rankings found via Scott-Knott are colored alternatively gray or white. In that table:
\bi
\item Initial results are the ``baseline`` set shown at the bottom. Our convention is to report the  baseline  budget as the number of rows in that data sets.  
\item All other rows    have an  evaluation budget set    from {\em \{20,25,30\}}. 
\item Anything below the baseline is doing worse than the original data set. In Table~\ref{tab:model_summary}, that means that for this data set, EI\_GPM has failed to optimize this data set.
\item Top ranked results are shown at the top of the table and labeled rank ``0''.
\ei

\subsection{Experimental Rig}\label{rig}

\begin{figure}[!t]
\caption{
Average optimizations seen across all data sets. 
Most   improvement were
after a few dozens samples 
Let {\em b4.mu} and {\em b4.lo}   be the mean and smallest Chebyshev distances seen in the   original data.
Let {\em now.mu} and  {\em now.sd }be the mean and standard deviation of the best Chebyshevs  seen in  20 repeats of our active learning experiments-(in this case, LLM warm starts followed by exploit). 
The blue plot shows {\em (now.mu - b4.lo) / (b4.mu - b4.lo)}; i.e. the improvement seen by optimization, normalized by the maximum possible improvement
(and for the blue line in this plot,  {\em lower}   values are 
{\em better}).
  }\label{diff}
\begin{center}
\includegraphics[width=4.5in]{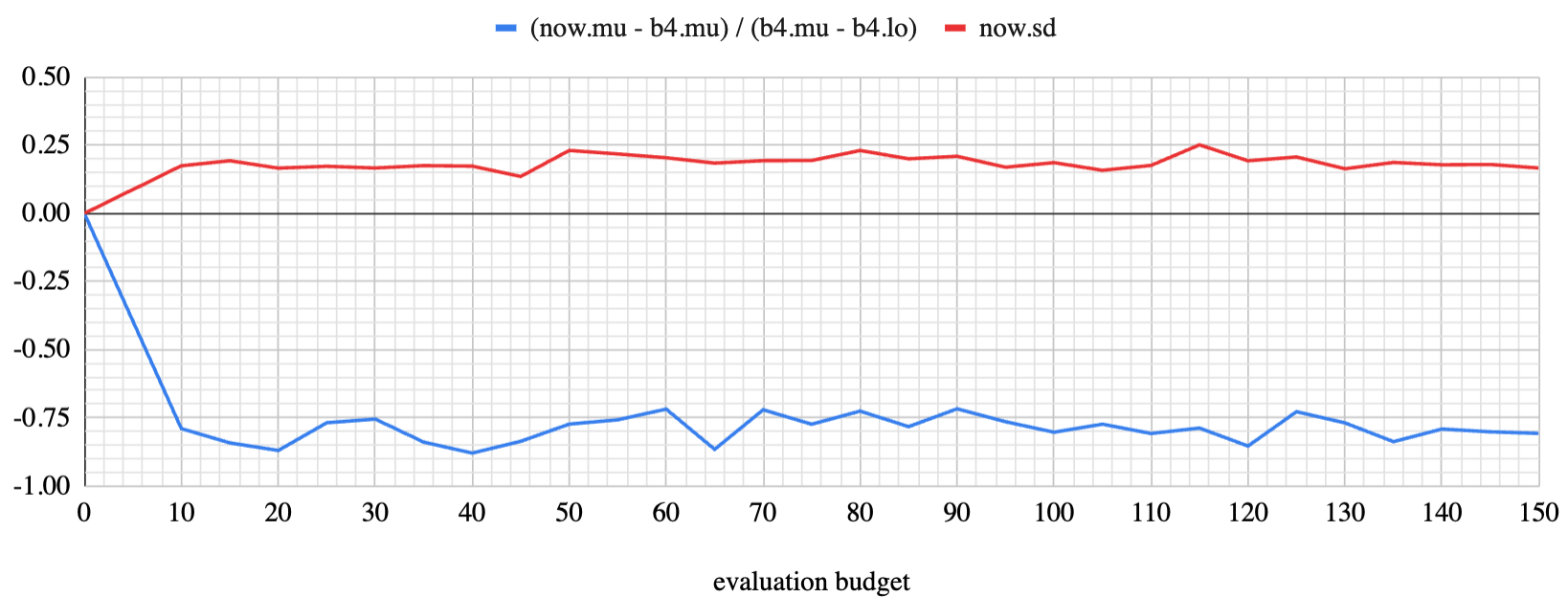}
\end{center}
\end{figure}

We ran all our active learners with $B_0$ evaluations for the warm starts and $B_1$ total evaluations where $B_0=4$ and $B_1 \in \{10,15,20,25,30\}$.
This was repeated 20 times for statistical validity.  
Budgets up to 30 were chosen since:
\bi
\item It fits the constraint of \S\ref{slowdata};
\item There are many results in the recent active learning literature
were the knee in the performance results occurs near  15 evaluations (e.g. see Figures 4,5,7 of~\cite{bilal2020best}).
\item In our active learning experiments,  most of the improvement is witnesses with $B_1 \le 30$. This ``30 is enough'' effect is seen in 
both
(a)~the   visualization of  Figure~\ref{diff};
and (b)~ a detailed statistical analysis. That statistical analysis 
built one plot like Figure~\ref{diff} for each data set. Then, for each plot,   the  
statistical methods of \S\ref{stats} search for better optimizations occurred for $B_1 > 30$.  None were found. 
\ei
To summarize all 49 data sets, we  generated one table like  Table~\ref{tab:model_summary}
for each data set. Next, we report the percentage of times a treatment appears at different ranks.
For example, Table~\ref{lo} shows that in 100\% of our low dimensional data sets, LLM/Exploit achieves
top rank.

\begin{table}[!b]
\centering
\scriptsize
\begin{tabular}{cc|ccccccc}
\hline
&\multicolumn{8}{c}{\textbf{Scott-Knott Rankings}}\\
\hline
Start & Acquire & 0 & 1 & 2 & 3 & 4 & 5 & 6 \\
\hline
LLM & Exploit  & \cellcolor{darkblue}\textbf{100} &  & &  &  &  &  \\
random & UCB\_GPM & \cellcolor{blue}\textbf{45} & 9 & 9 & & \cellcolor{lightblue}18 & 9& 9 \\
random & EI\_GPM        & \cellcolor{blue}\textbf{45} & \cellcolor{lightblue}18 & \cellcolor{lightblue}18 &  & 9&  & 9 \\
random & Exploit  & \cellcolor{blue}\textbf{27} & 9 & \cellcolor{lightblue}18 & \cellcolor{blue}27 &  9 & & 9 \\
random & PI\_GPM    & 9 & \cellcolor{blue}36 & \cellcolor{lightblue}18 & \cellcolor{lightblue}18 & 9 & &  \\

LLM & Explore   &   9 & 9 & \cellcolor{lightblue}18 & \cellcolor{lightblue}18 & 9& 9& 9 \\
& random    & \cellcolor{blue}36 & 9 &  & 9 & 9& 9 \\
random & Explore  & & \cellcolor{lightblue}18& & 9& \cellcolor{blue}27 & 9 &  \\
& Baseline  & & & & & 9 & \cellcolor{lightblue}18 &  \\
\hline
\end{tabular}
\caption{Percent frequency for some treatment appearing at some rank. e.g. ``50'' means that a treatment achieved a rank in half of our 49 data sets.  Results from 12 low dimensional  data sets (i.e. with   $x< 6$  independent features).}\label{lo}
\end{table}

\begin{table}[ht]
\centering
\scriptsize
\begin{tabular}{ll|ccccccc}
\hline
&\multicolumn{8}{c}{\textbf{Scott-Knott Rankings}}\\
\hline
Start & Acquire & 0 & 1 & 2 & 3 & 4 & 5 & 6 \\
\hline
LLM & Exploit  & \cellcolor{darkblue}\textbf{50} &  & 7 & 14 & 7 & 7 &  \\
random & UCB\_GPM    & \cellcolor{darkblue}\textbf{36} & \cellcolor{darkblue}36 & \cellcolor{lightblue}14 & \cellcolor{lightblue}14 &  & &  \\
random & PI\_GPM & \cellcolor{darkblue}\textbf{36} & \cellcolor{blue}14 & 14 & \cellcolor{lightblue}29 & & 7&  \\
random & EI\_GPM    & \cellcolor{darkblue}\textbf{36} & \cellcolor{lightblue}21 & \cellcolor{blue}14 & 7 & 7& 14 &  \\
random & Exploit  & \cellcolor{blue}\textbf{21} &  &  & 7 &   & 7& 7 \\
LLM & Explore      & \cellcolor{lightblue}14 & 7 &  &   &21 & 7& 7 \\
& random   & \cellcolor{blue}14 & 7 &  &  &  & 14 & \cellcolor{lightblue}21 \\
random & Explore  &14 & 7& & &  & 7 & 7 \\
& Baseline  &7 & & 7& &  &  &  \\
\hline
\end{tabular}
\caption{Frequency of ranks achieved.
Same format as Table~\ref{lo}.  For 16 medium dimensional   datasets (with $5< x < 11$   features).} \label{med}
\end{table}

\begin{table}[ht]
\centering
\scriptsize
\begin{tabular}{ll|ccccccc}
\hline
&\multicolumn{8}{c}{\textbf{Scott-Knott Rankings}}\\
\hline
Start & Acquire & 0 & 1 & 2 & 3 & 4 & 5 & 6 \\
\hline
random & UCB\_GPM & \cellcolor{darkblue}\textbf{50} & \cellcolor{blue}22 &  & \cellcolor{lightblue}17 & 6& &  \\
random & EI\_GPM & \cellcolor{darkblue}\textbf{44} &  & \cellcolor{blue}28 & 11 & 6& 6 &  \\
random & PI\_GPM    & \cellcolor{blue}\textbf{39} & \cellcolor{blue}28 & \cellcolor{lightblue}11 & \cellcolor{lightblue}11 & 6 & &  \\
LLM & Exploit  & \cellcolor{darkblue}\textbf{33} & 6 & \cellcolor{lightblue}22 & 6 & 6 & 6 & 6 \\

random & Exploit  & \cellcolor{blue}\textbf{17} & 6 & 6 &  &  \cellcolor{lightblue}17 & 11& 6 \\
LLM & Explore      & \cellcolor{lightblue}17 &   11 & 11& 6& \cellcolor{lightblue}17& 11 \\
& random   & \cellcolor{blue}11 & 6& 6 & 6 & 6 & 6 & 6 \\
random & Explore  & & 11& 6& &  & \cellcolor{lightblue}17 & 17 \\
& Baseline  & & & 11& 6&  & 6 &  \\
\hline
\end{tabular}
\caption{Frequency of ranks achieved.
Same format as Table~\ref{lo}.  For 19 high dimensional   datasets (with $x>10$ independent features).}\label{hi}
\end{table}

\newpage \section{Results for Research Questions}
\label{Results}

\subsection{{\bf RQ1}: {\em Is   active learning useful for SE tasks?} }

\begin{wraptable}{r}{2.6in}
\scriptsize
\centering
\begin{tabular}{ll|ccccccc}
\hline
&\multicolumn{8}{c}{\textbf{Skott-Knott Rankings}}\\
\hline
Start & Acquire & 0 & 1 & 2 & 3 & 4 & 5 & 6 \\
\hline
random & UCB\_GPM & \cellcolor{lightred}19 & 21 & 22 & 21 & 22&15 & 15 \\
random & EI\_GPM & \cellcolor{lightred}19 & 17 & 23 & 23 & 18& 18 &  18\\
random &PI\_GPM    & \cellcolor{red}17 & 18 & 22 & 21 & 15 & 15&  \\
LLM & Exploit  & \cellcolor{red}17 & 15 & 18 & 18 & 18 & 20 & 25 \\

random & Exploit  & \cellcolor{lightred}19 & 23 & 15 &  19&  21 & 17& 20 \\
LLM & Explore      & \cellcolor{darkred}16 &   15 & 18& 18& 17& 16 \\
& random   & \cellcolor{red}18 & 19& 20 & 20 & 18 & 19 & 19 \\
random & Explore  &22 & 19& 15& 15 & 22 & 18 & 21\\
\hline
\end{tabular}
\caption{Scott-Knott rankings comparison for number of evaluations needed}
\label{budget}
\end{wraptable}

The point of active learning is that AI tools should be able to learn 
a domain, after seeing very  few  labeled examples. Hence,
one way to assess active learning is to ask ``how any labels does it require to find good optimizations of the MOOT data?''

As illustrated in Figure~\ref{diff}, the steep decline in the blue curve (representing normalized improvements in performance) shows that most of the optimization benefit   is achieved with fewer than 30 labeled samples.
Also, 
Table~\ref{budget} shows,  on average, how many evaluations were required for a treatment to achieve a particular rank (averaged across all data sets).  As shown in the rank ``0'' column of that table,   rank0 was achieved after 16 to 22 evaluations.

 Another way to assess active learning is to ask ``how good is it versus just performing random selection''? The rows
with  ``Acquire=random'' in the Tables~\ref{lo},\ref{med},\ref{hi} show the results of just selecting  {\em \{10,15,20,25,30\}} rows at random, sorting them by their Chebshev distance, then returning the best one.
In no case in  
Table~\ref{lo}, \ref{med} or \ref{lo} did a purely
random method score appear most often in the rank ``0'' column.



In summary, we say:

\begin{formal}{\bf RQ1:} 
For  the SE tasks studied here, active learning is useful
since it can explore thousands to tens of thousands of rows of data after just a few dozen labeling. 
\end{formal}

\subsection{{\bf RQ2}: {\em Are   warm-starts   useful for active learning?} }

A frequently asked question of this work is ``do warm starts with only $B_0=4$ examples improve optimization results?''. 
To answer that question, we can look at our results:
\bi
\item
For low  dimensionality data sets,  in Table~\ref{lo}
we see that (a)~LLM/Exploit earns top rank 100\% of the time
while (b)~random/Exploit only earns top rank 27\% of the time.
\item
For medium  dimensionality data sets,  in Table~\ref{med}
we see that  (a)~LLM/Exploit earns top rank 50\% of the time
while (b)~random/Exploit only earns top rank 21\% of the time.
\ei
Hence we say:
\begin{formal}{\bf RQ2:} 
For the SE tasks studied here, even
if warm starts select a handful of initial examples,
then those choices can still result in major improvements
to optimization results.
\end{formal}

\subsection{{\bf RQ3}: {\em Are LLMs the best way to generate warm starts?} }\label{rq3}

The success of LLM's warm starts are   remarkable, especially  considering   what we  were asking LLMs to do. Recalling 
Table~\ref{fig:enter-label}, our experiments
ask    LLMs to invent
two better rows and two worse rows than an initial set of  $B_0=4$ rows.
 Geometrically, this means that LLMs had to:
 \bi
 \item
 Perform a multi-dimensional, almost PCA-like analysis\footnote{Principal Component Analysis (PCA) is a dimensionality reduction technique that transforms data into a set of uncorrelated variables called principal components, which capture the maximum variance. It simplifies complex datasets by projecting them onto fewer dimensions while preserving essential patterns.}  to find the dimensions that matter then most, 
 \item 
 Then perform  an extrapolation along those dimensions.
 \ei
  To be sure, that approach failed for larger dimensional problems.
However, 
 the fact that this performed so well for low and medium dimensional examples
 is a testament to the inherent  power of these large languages.

One caveat to the last paragraph is that there are some nuances
that need to be discussed on the effectiveness of LLMs for warm starts. Observe that LLM/exploit performed very well in Table~\ref{med}
and better than anything else is Table~\ref{lo}. However, LLM paired with other acquisition functions did not perform well. Specifically, LLM/Explore performed much worse than LLM/Exploit.
This results suggests that LLMs has thoroughly searched  through the zone of doubt described  \S\ref{TPE}, thus making further exploration
superfluous. Once again, this result speaks to the power of LLM-based inference.

All that said, LLM/exploit failed for higher-dimensional problems, at which point randomly selected starts followed by UCB\_GPM performed best.
Clearly, further research is required in order to extend LLM-based reasoning to more complex tasks.

Overall, we summarize these results as follows:

\begin{formal}{\bf RQ3:}
For the SE tasks studied here, LLM-based warm starts following by exploit-guided acquisition performs best for  low and medium dimensional problems. However, LLM effectiveness diminishes in high-dimensional problems, where Bayesian methods like
Gaussian Process Models perform best
\end{formal}

\section{Discussion}

\subsection{Why Did LLMs Fail for Higher-dimensional Data?}

 We   applaud the success (reported above) of LLMs for low and medium complexity problems. But why does this approach fail for   higher-dimensional data?

Perhaps the answer lies in the data used to train LLMs. 
 Large language models train from data    available ``in the commons''; i.e. all the data generated by (say) programmers who
store their code in Github. In    the commons, there may exist   many acceptable   solutions
for (e.g.) how to build a website in Python. Given a plethora of such solutions,  LLMs can offer a useful     response  to a specific prompt.

Outside the commons,  there are problems 
that
  humans rarely address or, if they do, they rarely produce solution that a broad community would find acceptable.  
For such ``uncommon'' tasks, LLMs may lack   sufficient training data.
Many of the optimization tasks in MOOT are ``uncommon''. 
For example:
\bi
\item 
Our XOMO* data sets come from books discussing process options for software projects. These data sets list 24 parameters, usually discretized into five ranges
(very low, low, nominal, high, very high).  While many publications mention these choices, we know of none that conclude that one of these  $5^{24}\approx10^{16}$ choices is undeniably better
than the rest. 
\item Several of our models refer to the configuration of cloud-based software systems. We would also call  this an ``uncommon'' problem since there are some few publicly available examples of well-configured cloud 
environments\footnote{See Table 7 of Tang et al.~\cite{Tang23a} 
for a long list of cloud configuration errors. Also see industrial reports such as the 2024 Verizon business data breach report that states   80\% of all security breeches on the cloud are configuration-related \url{https://www.verizon.com/business/resources/reports/dbir/}.}.
\ei

\begin{wrapfigure}{r}{2.5in}
\begin{center}
\includegraphics[width=2.5in]{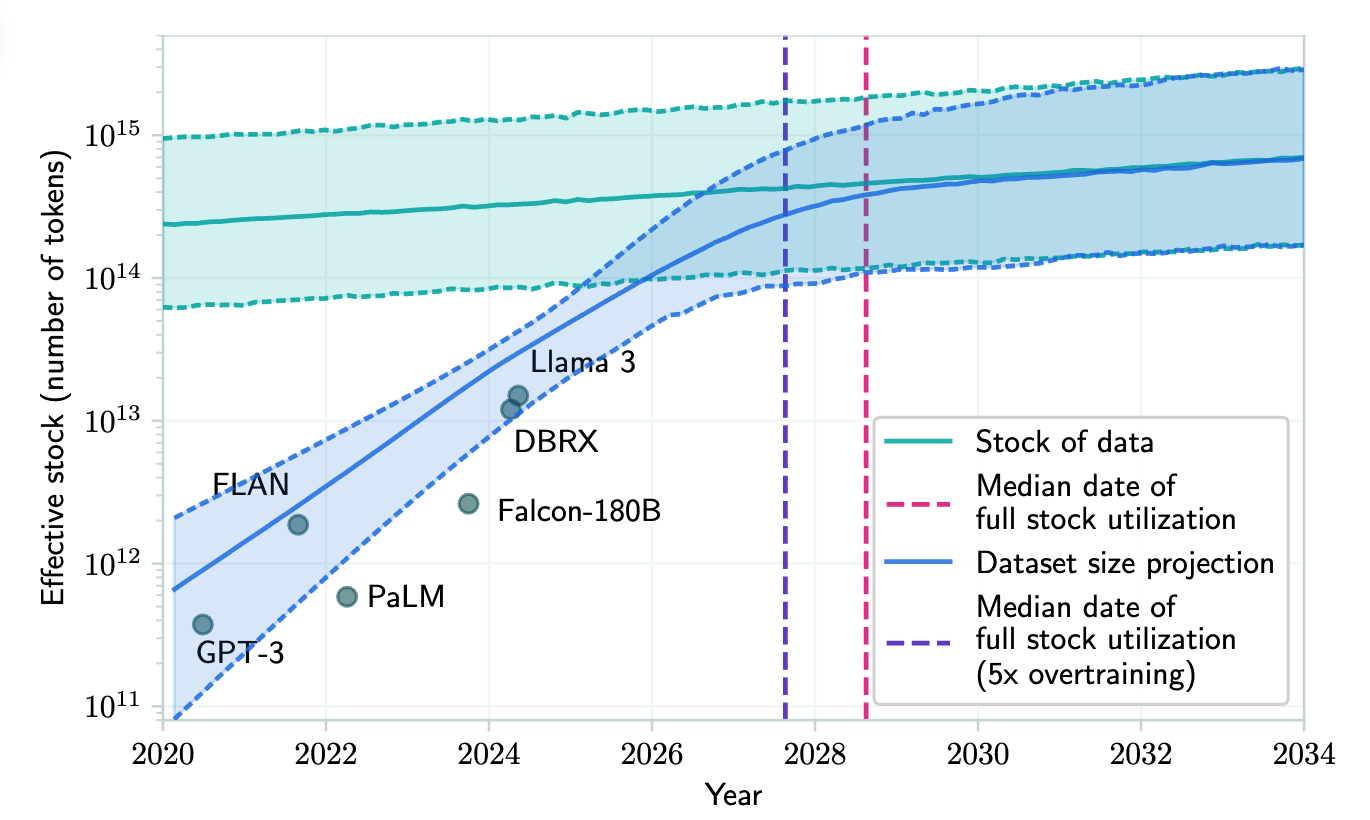}
\end{center}
\caption{Median results from
the Villalobos et al. model (shown in green) estimate that  by 2028, we will run out of new textual data needed to train    bigger and better LLMs~\cite{villalobosposition}.}\label{fig:data1}
\end{wrapfigure} 
We therefore offer the following conjecture.
For low- and medium- dimensional   SE  problems LLMs can  make  find effective initial candidates, even for uncommon problems. 
 The same is not true for higher-dimensional problems, an observation we attribute to the nature of uncommon problems:
\bi
\item
The more complex the problem ...
\item
... and the fewer examples of good solutions in the commons ...
\item
... then the harder it becomes for LLMs to learn a model that works for that problem.
\ei
For these harder uncommon problems, we will recommend other methods (GPM or TPE).

Looking into the future, we predict that, increasingly, LLMs will be challenged by ``uncommon'' problems.
Recent results strongly suggest
that LLMs may stop improving, very soon.
 In 2024, Villalobos et al. 
noted that new LLMs require exponentially increasing amounts of training data~\cite{villalobosposition}. In Figure~\ref{fig:data1}, they project that newer and larger models will soon exhaust the available textual data\footnote{One limitation of the  Villalobos et al. analysis is that it is only based on text tokens are there are other media that could be explored (e.g. visual). Nevertheless, their general point remains. Learning processes
that require exponentially more data will soon exhaust the available training data.}.  At that time, research work like this paper will be required to handle all the problems for which LLMs lack sufficient training data.

(Aside: Villlobios et al. warn that methods to extend this data (e.g. using one LLM to generate new data to train another) can actually degrade performance since automatically generated data may lacks the diversity needed for good inference~\cite{singh2023beyond}.)

\subsection{Threats to Validity}

Despite the promising results, this study is subject to several threats to validity that must be carefully considered to ensure accurate interpretation of the findings.

\paragraph{Internal Validity.}  
A primary concern is the potential for implementation errors in the active learning algorithms and acquisition functions. Although rigorous validation steps were taken, undetected bugs or implicit assumptions could skew results. Moreover, the selection of hyperparameters for Gaussian Process Models (GPM) and Tree of Parzen Estimators (TPE) may introduce subtle biases. To mitigate this, sensitivity analyses and replication by external researchers are essential.

\paragraph{External Validity.}  
While the datasets span multiple SE tasks, they certainly do   encapsulate the diversity of real-world SE problems. As a result, the findings may not generalize to tasks involving vastly different domains, such as next-release planning or SE reinforcement learning. Future work should expand the dataset to encompass a broader variety of SE optimization problems to improve generalizability.

\paragraph{Construct Validity.}  
This study relies on Chebyshev distance to evaluate solution quality, which, while standard in multi-objective optimization, may overlook nuanced trade-offs inherent to SE tasks. Exploring alternative or composite metrics could reveal additional insights and better reflect the complexities of SE decision-making processes.

\paragraph{Conclusion Validity.}  
The use of Scott-Knott clustering and Cliff's Delta provides a statistically grounded analysis. However, the small sample size of labeled data points introduces the risk of both type I and type II errors. Expanding the experimental budget and incorporating complementary statistical methods would enhance the robustness of the conclusions.

\paragraph{Learner Bias}
We have used Google Gemini 1.5 pro as our model for generating synthetic warm starts, The kind model used can significantly affect the quality of the generated samples and it is impossible to compare the performances of various models due to restrictions.

\subsection{Future Research  }

Building on the core research questions outlined in this study, several avenues for further exploration emerge that can enhance the understanding of LLM-based warm starts in software engineering (SE) active learning tasks.

\paragraph{Generalization and Transferability.} As
discussed in our literature review,
most research papers in this arena test their methods on less than half a dozen data sets. Here, we have explored 49 so this work is less susceptible that most papers
to criticisms of ``lack of generality''. That said, it would be wise to continue the testing of this method. Whenever new multi-objective SE data becomes available, we plan to apply the techniques of this paper to that data.  

\paragraph{Dimensionality and Complexity.}  
Given the observed decline in LLM effectiveness for high-dimensional problems, future research could explore whether dimensionality reduction techniques (e.g., PCA) can mitigate this limitation.

\paragraph{Long-Term Performance and Iteration.}  
Future studies may also examine how LLM-generated warm starts evolve over multiple iterations of active learning. For example, can results from previous cycles further improve future warm starts?

\begin{wrapfigure}{r}{3in}
\includegraphics[width=3in]{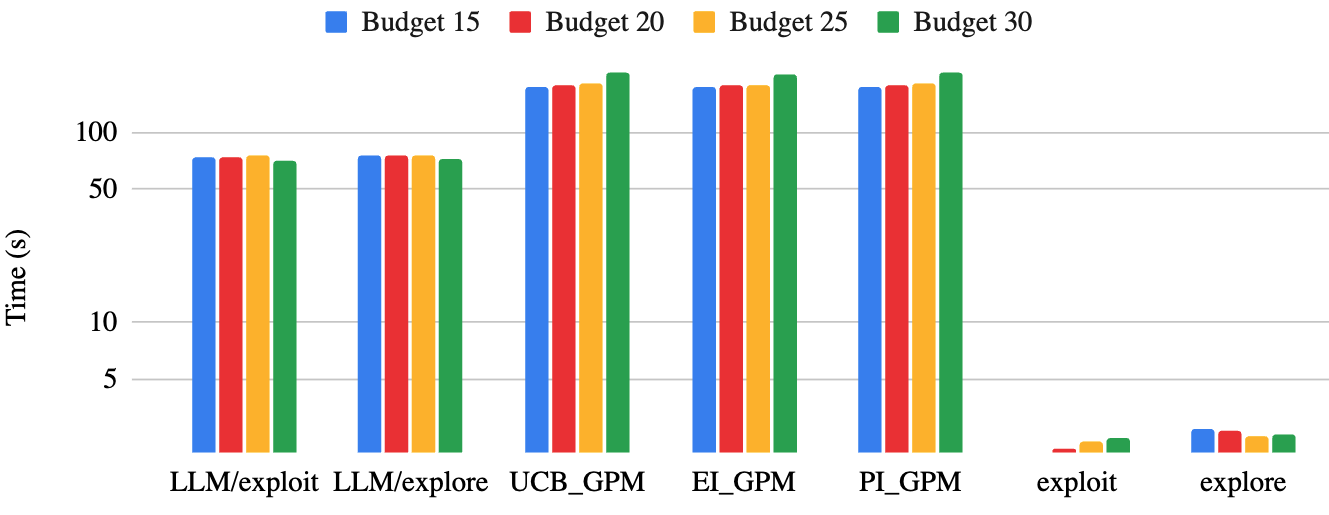}
\caption{A sample of runtimes from these experiments.}\label{runtimes}
\end{wrapfigure} 
\paragraph{Cost and Efficiency.}  
Another critical line of inquiry is the computational cost of using LLMs for warm starts compared to Bayesian methods. 
Usually, the runtime costs of LLMs are the most prohibitively expensive item in an SE optimization study. Interestingly, here, that it not the case. Gaussian Process Models compute the   mean and standard deviation of  estimates by running the available data across a wide range of possible kernels. As shown in Figure~\ref{runtimes}, this process scan be slower even than LLM few shot learning. 

One other thing to note from Figure~\ref{runtimes} is that the TPE methods (explore and exploit) run very much faster than GPM or LLM-based methods. Future research could explore if initial quick ``peeks'' at the data (with TPE) could inform and improve subsequent reasoning with LLMs or GPM.

\paragraph{Interpretability and Explainability.}  
Given the low number of evaluations used in these studies
($B_1 \le 30$), then explanations generated from active learning for LLMs could be very simple indeed. For example, perhaps there is some way to use active learning as a post-processor to  LLMs   to find a small number of easily explain examples.  learning.

\section{Conclusion}

This study demonstrates that Large Language Models (LLMs) can effectively warm-start active learning for software engineering (SE) tasks, significantly improving performance for low- and medium-dimensional problems. By generating plausible initial guesses, LLMs reduce the labeling effort required, outperforming random starts and matching or exceeding Bayesian methods in many cases.  

However, for high-dimensional tasks, LLMs face limitations, with Gaussian Process Models (GPM) continuing to deliver superior results. This highlights the complementary roles of LLMs and traditional Bayesian approaches in SE optimization.  

Our  contributions include:  
\begin{itemize}  
    \item A novel application of LLMs for warm-starting SE active learners.  
    \item Empirical validation across 49 multi-objective SE tasks.  
    \item Open-source release of data and scripts to promote reproducibility.  
\end{itemize}  

Future work will focus on extending LLM effectiveness to higher dimensions, exploring hybrid models, and validating results on broader SE tasks. This study underscores the potential of LLMs to enhance optimization in data-scarce SE environments.

\bibliographystyle{ACM-Reference-Format}
\bibliography{sample-base,timm}

\end{document}